# Machine Learning versus Mathematical Model to Estimate the Transverse Shear Stress Distribution in a Rectangular Channel

Babak Lashkar-Ara [1,*], Niloofar Kalantari [1], Zohreh Sheikh Khozani [2] and Amir Mosavi [2,3,4,*]

1 Department of Civil Engineering, Jundi-Shapur University of Technology, 64616-18674 Dezful, Iran;
2 Institute of Structural Mechanics, Bauhaus Universität-Weimar, 99423 Weimar, Germany; zohreh.khozani.sheikh@uni-weimar.de
3 School of Economics and Business, Norwegian University of Life Sciences, 1430 Ås, Norway
4 John von Neumann Faculty of Informatics, Obuda University, 1034 Budapest, Hungary
* Correspondence: Lashkarara@jsu.ac.ir (B.L.-A.); amir.mosavi@uni-weimar.de (A.M.)

**Abstract:** One of the most important subjects of hydraulic engineering is the reliable estimation of the transverse distribution in the rectangular channel of bed and wall shear stresses. This study makes use of the Tsallis entropy, genetic programming (GP) and adaptive neuro-fuzzy inference system (ANFIS) methods to assess the shear stress distribution (SSD) in the rectangular channel. To evaluate the results of the Tsallis entropy, GP and ANFIS models, laboratory observations were used in which shear stress was measured using an optimized Preston tube. This is then used to measure the SSD in various aspect ratios in the rectangular channel. To investigate the shear stress percentage, 10 data series with a total of 112 different data for were used. The results of the sensitivity analysis show that the most influential parameter for the SSD in smooth rectangular channel is the dimensionless parameter $B/H$, Where the transverse coordinate is $B$, and the flow depth is $H$. With the parameters ($b/B$), ($B/H$) for the bed and ($z/H$), ($B/H$) for the wall as inputs, the modeling of the GP was better than the other one. Based on the analysis, it can be concluded that the use of GP and ANFIS algorithms is more effective in estimating shear stress in smooth rectangular channels than the Tsallis entropy-based equations.

## 1. Introduction

Knowledge of boundary shear stress is necessary when studying sediment transport, flow pattern around structures, estimation of scour depth and channel migration. The determination of boundary shear stress, i.e., at the wall and bed depends on the channel geometry and its associated roughness. Various direct and indirect methods have been extensively discussed in experimentally measure the wall and bed shear stresses in channels with different cross sections [1–4]. Bed shear stress can be estimated based on four techniques (1) bed slope product $\tau_b = gHS$ , (2) law of the wall velocity profiles $\frac{u}{u_*} = \frac{1}{k} ln\left(\frac{z}{z_0}\right)$ where $\tau_b = \rho u_*^2$, (3) Reynolds stress measurement $\tau_b = \rho(-\overline{u'w'})$ and (4) turbulent kinetic energy (TKE), $TKE = \frac{\rho\left(\overline{u'^2}+\overline{v'^2}+\overline{w'^2}\right)}{2}$, where $\tau_b = C_1 TKE$ , where $u'$, $v'$ and $w'$ are the fluctuating horizontal, transversal and vertical velocity components, respectively and $C_1 = 0.20$ [5]. The symbols $g$, $H$ and $S$ denote gravity, water level and channel slope, respectively, whereas $u$ is the velocity at height $z$, $u_*$ is the shear velocity, $k$ is von Karman constant and $z_0$ is the roughness length.

These methods are useful in presenting a point-based representation of shear stress in a channel, whereas the shear stress distribution (SSD) provides a more accurate hydrodynamic profile within a channel. Knight and Sterling [6] measured the SSD in a circular channel with and without sediment. They examined a wide range of flow depths for each level benching and therefore it had been possible to determine the extent to which the hydraulics changes Park et al. [7] utilized laboratory-scale water flume and measured the bed shear stress under high-velocity flow conditions directly. Lashkar-Ara and Fatahi [8] measured transverse SSD in the channel bed and wall by using an optimal diameter Preston tube to evaluate the SSD on a rectangular open channel. The outcome of this research is two-dimensional relationships to evaluate local shear stress in both bed and wall. The bed and wall relative coordinates $b/B$ and $z/H$ in the cross section also the aspect ratio $B/H$ are the function of these relationships. The study showed that the dimensionless SSD is greatly affected by the aspect ratio. Utilizing the advantages offered in the soft computing method and the artificial intelligence (AI) techniques, other researchers have been extended numerically and analytically to overcome difficulties with experimental measurements [9–12]. Martinez-Vazquez and Sharifi [13] utilized recurrence plot (RP) analysis and eigenface for recognition to estimate the SSD in trapezoidal and circular channels. A new approach has been developed by Sterling and Knight [14] to estimate the SSD in a circular open channel. In terms of accuracy, the analysis showed that there is a lack of ability in the outcome and it is not satisfactory. The uncertainty of the estimation of the model parameters and the high sensitivity of the outcomes to the expected experiment parameters can be due to this. Sheikh Khozani and Bonakdari [15] extended the analytical method based Renyi entropy to estimate SSD in circular channels. Sheikh Khozani and Bonakdari [16] researched on the comparison of five different models in straight compound channel prediction of SSD. In other research, Sheikh Khozani and Wan Mohtar [10] analyzed the formulation of the SSD on the basis of the Tsallis entropy in circular and trapezoidal channels. Sheikh Khozani et al. [17] have attempted in another study to use an improved SVM method to estimate shear stress in rough rectangular channel.

Ardiçlioğlu et al. [18], conducted an experimental study for the SSD throughout the entire length of the cross-section in fully developed boundary layer area, in an open rectangular channel, in both smooth and rough surface. By measuring the speed in both smooth and rough surfaces, they conducted tests. Using logarithmic distribution of velocity, the average shear stresses in the cross section for aspect ratios of 4.2–21.6 and the Froude numbers of 0.12–1.23 were measured. The definition of the Tsallis entropy was used by Bonakdari et al. [19] to predict the SSD in trapezoidal and circular channels and achieve acceptable accuracy. Although the direct measurement of shear stress in laboratory provides correct description of the spatial pattern, the measurement of shear stress using shear place or cell is laborious, complex, requires careful calibration and may not applicable to all type of channels [20]. The use of soft computing techniques in the simulation of engineering problems was intensively studied and a variety of soft computing methods were suggested. To approximate the daily suspended sediment load, Kisi et al. [21] used a genetic programming (GP) model. They also contrasted this approach with various machine learning methods and concluded that the GP model works better than the others. In estimating SSD in circular channels with and without flat-bed Sheikh Khozani et al. [22,23] applied randomize neural network (RNN) and gene expression programming (GEP). In this study, the Tsallis entropy was used to determine SSD in a smooth bed and wall in a rectangular open channel. This is then used to measure the SSD in various aspect ratios in the rectangular channel. In the second part of the study, two soft computing methods were applied to predict the transverse of SSD in the smooth rectangular channel. The methods of genetic programming (GP) and the adaptive neuro-fuzzy inference system (ANFIS) were examined to determine the precision of these models in estimating bed and wall shear stress. This study aimed at using the Tsallis entropy method to predict the SSD in the smooth rectangular channel. The results of the Tsallis entropy, GP and ANFIS methods compared with experimental results of Lashkar-Ara and Fatahi [8]. Although this analysis was performed in parallel with Sheikh Khozani and Bonakdari [16] research, it can be said in a practical contrast that the data used in this study is based on the measurement of shear stress using the optimal diameter of the Preston tube, which was designed by Lashkar-Ara and Fatahi [8], so the comparison of findings is more precise and less uncertain.

## 2. Materials and Methods

### 2.1. Data Collection

Information on the SSD was collected in the Lashkar-Ara and Fatahi [8] experiments of a smooth rectangular channel, performed in a flume 10-meter long, 60 cm wide and 70 cm high. All measurements were performed in the range of 11.06–102.38 liter per second flow rate. Flow rate variations led to observable changes in water depth ranging from 4.3 to 21 cm and the aspect ratio of 2.86–13.95. The values of static and total pressure difference in various aspect ratios of $B/H$ were measured and reported using pressure transducer apparatus with a capacity of 200 mill bar and 50 Hz measuring frequency. In order to create uniform flow condition and to match the hydraulic gradient with the flume bed slope a weir at the end of the flume was installed. Figure 1 illustrates the notation used for a smooth rectangular channel conduit. Figure 2 shows the schematic of experimental setup.

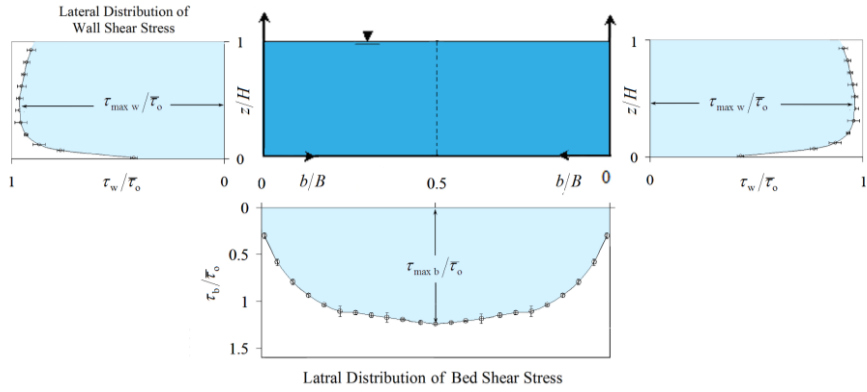

**Figure 1.** Schematics of local shear stress distribution coordinates in the rectangular channel wall and bed.

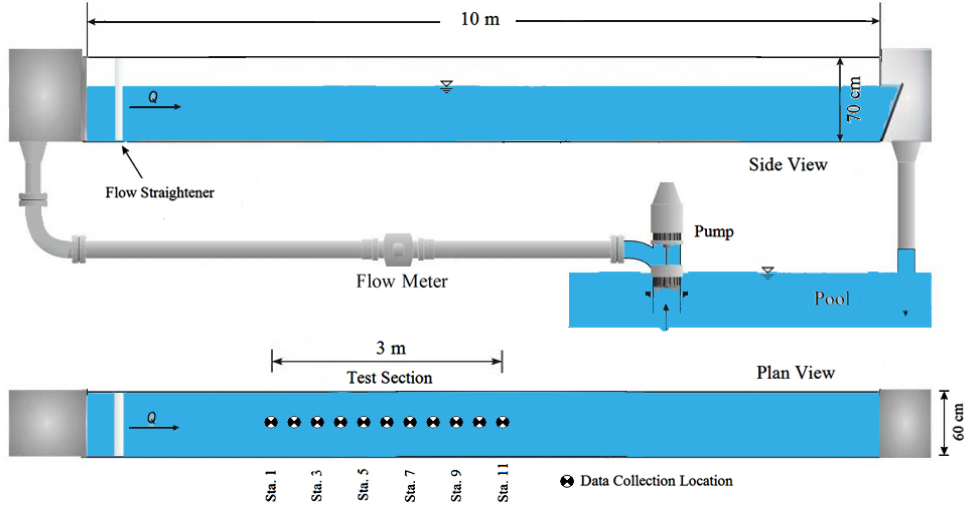

**Figure 2.** Experiment schematic.

Based on previous studies in the laboratory and field investigation, the effective criteria for evaluating the SSD along the wet periphery of a channel can be expressed as follows:

$$f_1(\bar{\tau}_w, \rho, \upsilon, g, V, H, S_w, S_o, B, z, K_s) = 0 \tag{1}$$

$$f_2(\bar{\tau}_b, \rho, \upsilon, g, V, H, S_w, S_o, B, b, K_s) = 0 \tag{2}$$

where $\bar{\tau}_w$ is the average wall shear stress, $\bar{\tau}_b$ is the average bed shear stress, $\rho$ is the density, $\upsilon$ is the kinematic viscosity, $g$ is the gravity acceleration, $V$ is the flow velocity, $H$ is the flow depth, $B$ is the flume bed width, $S_w$ is the water surface slope, $k_s$ is the roughness height, (Re) is the Reynolds number and (Fr) is the Froude number.

The Buckingham-π theorem is used to obtain independent dimensional parameters for wall and bed shear stress, as shown in Equations (3) and (4).

$$f_3\left(\frac{\upsilon}{VH}, \frac{K_s}{H}, \frac{gH}{V^2}, \frac{B}{H}, \frac{z}{H}, \frac{\overline{\tau}_w}{\rho g H S_w}\right) = 0 \quad (3)$$

$$f_4\left(\frac{\upsilon}{VH}, \frac{K_s}{H}, \frac{gH}{V^2}, \frac{B}{H}, \frac{b}{B}, \frac{\overline{\tau}_b}{\rho g H S_w}\right) = 0 \quad (4)$$

In the case of smooth channel equations (3) and (4) can be rewritten as (5) and (6):

$$\frac{\overline{\tau}_w}{\rho g H S_w} = f_5\left(\text{Re}, \text{Fr}^2, \frac{B}{H}, \frac{z}{H}\right) \quad (5)$$

$$\frac{\overline{\tau}_b}{\rho g H S_w} = f_6\left(\text{Re}, \text{Fr}^2, \frac{B}{H}, \frac{b}{B}\right) \quad (6)$$

For GP simulation, 160 data of bed shear stress ($\tau_b$) and 100 data of wall shear stress ($\tau_w$) were collected in a smooth rectangular channel with different flow depths. Approximately 70 percent of the total data were chosen for training and the remaining 30 percent for testing. The summary of experiments is tabulated in Table 1.

**Table 1.** Experimental summary.

| Parameters | Variable Definition | Minimum | Maximum | Mean |
|---|---|---|---|---|
| $H$ (m) | Flow depth | 0.043 | 0.21 | 0.0928 |
| $B/H$ | aspect ration | 2.86 | 13.95 | 7.98 |
| $Q$ (L/s) | Discharge | 11.06 | 102.38 | 34.795 |
| $V$ (m/s) | Velocity | 0.429 | 0.813 | 0.568 |
| Fr | Froude number | 0.66 | 0.566 | 0.618 |
| Re $\times 10^4$ | Reynolds number | 6.4 | 39.87 | 16.418 |
| Re$_*$ | Shear Reynolds | 0.322 | 0.609 | 0.426 |
| $\gamma HS$ | Total shear stress | 0.442 | 2.162 | 0.955 |

*2.2. Tsallis Entropy*

If a random variable ($\tau$) in a cross section of a channel is assumed to be a consistent shear stress, then, according to Tsallis entropy of [24] the SSD or shear stress probability density function f($\tau$), can be identified as [19]:

$$H(\tau) = \frac{1}{q-1}\int_0^{\tau_{max}} f(\tau)(1 - f(\tau)^{q-1})\, d\tau \quad (7)$$

where $\tau$ is the shear stress, $q$ is a true number, and Tsallis's entropy function is $H(\tau)$. The $\tau$ value varies from 0 to $\tau_{max}$, and with these restrictions, the integral value of $H(\tau)$ will be 1.

Using the maximum entropy theorem, the f($\tau$) can be calculated to maximize the entropy function subject to specified constraints like Equations (8) and (9) respectively [25].

$$\int_0^{\tau_{max}} f(\tau)d\tau = 1 \quad (8)$$

$$\int_0^{\tau_{max}} \tau.f(\tau)d\tau = \overline{\tau} \quad (9)$$

where the mean and maximum shear stress values are $\overline{\tau}$ and $\tau_{max}$, respectively.

At this stage, using maximization of Lagrange coefficients by Equations (7)–(9), the Lagrange function L can be written down as Equation (10):

$$L = \int_0^{\tau_{max}} \frac{f(\tau)}{q-1}(1-f(\tau)^{q-1})d\tau + \lambda_0\left(\int_0^{\tau_{max}} f(\tau)d\tau - 1\right) + \lambda_1\left(\int_0^{\tau_{max}} \tau.f(\tau)d\tau - \bar{\tau}\right) \quad (10)$$

where $\lambda_0$ and $\lambda_1$ are the Lagrange multipliers. By $\partial L/\partial(\tau) = 0$ to maximize entropy, the $f(\tau)$ yields as:

$$f(\tau) = \left[\frac{q-1}{q}(\lambda' + \lambda_1.\tau)\right]^{\frac{1}{(q-1)}} \quad (11)$$

where $\lambda' = \frac{1}{q-1} + \lambda_0$. In Equation (10), the shear stress probability distribution function (PDF) is represented by $f(\tau)$. The SSD's cumulative distribution function (CDF) is introduced as Equation (12):

$$F(\tau) = \int_0^{\tau_{max}} f(\tau)d\tau = \frac{y}{L} \quad (12)$$

where $y$ is the direction of the channel wall, which varies from 0 at the free surface to $L$, and $L$ is the entire wetted perimeter. The function of $f(\tau)$ is the derivative of $F(\tau)$, so a partial derivation of $F(\tau)$ with respect to y is carried out in the following equation:

$$f(\tau) = \frac{dF(\tau)}{du} = \frac{1}{L}\frac{dy}{d\tau} \quad (13)$$

By substituting Equation (11) into Equations (12) and (13) and solving the integral and simplifying, the shear stress function is represented as Equation (14).

$$\tau = \frac{k}{\lambda_1}\left[(\frac{\lambda'}{k})^k + \frac{\lambda_1 y}{L}\right]^{\frac{1}{k}} - \frac{\lambda'}{\lambda_1} \quad (14)$$

where k = q/q − 1 and q value is the constant of ¾ according to [10,26], which is defined as the parameter of the Tsallis relationship. $\lambda_1$ and $\lambda'$ are Lagrange multipliers that can be derived by trial and error from two implicit equations that follow. Indeed, by inserting and integrating Equation (10) into two constraints (Equations (8) and (9)), two Equations (15) and (16) are returned as:

$$[\lambda' + \lambda_1\tau_{max}]^k - [\lambda']^k = \lambda_1 k^k \quad (15)$$

$$\tau\lambda_1\left[\lambda' + \lambda_1\tau_{max}[]^k\left[\lambda' + \lambda_1\tau_{max}[]^{k+1}[\lambda']^{k+1}k\lambda_1^2 k^{-k}\right]\right]_{max} \quad (16)$$

Equations (15) and (16) solve to obtain two undefined Lagrange multipliers ($\lambda_1$ and $\lambda'$). To estimate the SSD, a pair of mean and maximum shear stresses is required. The results of the Lashkar-Ara and Fatahi [9] studies have been used for this reason in order to estimate the values of $\tau_{max}$ and $\bar{\tau}$. They adjusted the slope of the bed flume at $9.58 \times 10^{-4}$. The shear stress carried by the walls and bed was measured for a different aspect ratio (B/H = 2.86, 4.51, 5.31, 6.19, 7.14, 7.89, 8.96, 10.71, 12.24 and 13.95). For each aspect ratio, the distribution of shear stress in the bed and wall was measured by a Preston tube. The best fit equation was obtained for $\tau_{max}$ and $\bar{\tau}$ separately for wall and bed in aspect ratio 2.89 < B/H < 13.95 by assuming a fully turbulent and subcritical regime among all the experimental results. Relationships are shown in Equations (17)–(20).

$$\frac{\bar{\tau}_w}{\rho g RS} = \frac{2.1007 + 0.0462\left(\frac{B}{H}\right)}{1 + 0.1418\left(\frac{B}{H}\right) + \left(\frac{B}{H}\right)^{-0.0424}} \quad (17)$$

$$\frac{\bar{\tau}_b}{\rho g R S} = \frac{2.0732 - 0.0694\left(\frac{B}{H}\right)}{1 - 0.146\left(\frac{B}{H}\right) + \left(\frac{B}{H}\right)^{-0.1054}} \quad (18)$$

$$\frac{\tau_{max\ w}}{\rho g R S} = \frac{2.5462 + 6.5434\left(\frac{B}{H}\right)}{1 + 6.34\left(\frac{B}{H}\right) + \left(\frac{B}{H}\right)^{-0.1083}} \quad (19)$$

$$\frac{\tau_{max\ b}}{\rho g R S} = \frac{3.157 + 0.8214\left(\frac{B}{H}\right)}{1 + 0.8535\left(\frac{B}{H}\right) + \left(\frac{B}{H}\right)^{-0.1401}} \quad (20)$$

where $\bar{\tau}_w \& \bar{\tau}_b$ and $\tau_{max\ w}$ and $\tau_{max\ b}$ are the mean and maximum shear stress on the channel wall and bed, respectively. Therefore, the transverse SSD for the rectangular open channel can be determined depending on the aspect ratio and the slope of the channel bed.

### 2.3. Genetic Programming (GP)

In the second part of this analysis, the GP model is applied as one of the evolutionary algorithms (EA) to improve the accuracy of the given relations. The GP is an automated programming method to solve problems by designing computer programs GP is widely used for modeling structure recognition technology applications concerns. For this aim the GP technique was used to understand the basic structure of a natural or experimental process. The GP can optimize both the structure of the model and its parameters. One of the advantages of the GP algorithm is that it can extract an equation based the input and output parameters and it is more effective than other ANN models [27]. Table 2 represents the used parameters in modeling with GP algorithm including function set, the terminal set for $\frac{\tau_b}{\bar{\tau}_b}$, and the terminal set for $\frac{\tau_w}{\bar{\tau}_w}$. Further values of the parameters, i.e., number of inputs, the fitness function, error type, crossover rate, mutation rate, gene reproduction rate, population size, number of generations, tournament type, tournament size, max tree depth, max node per tree, and constants range can be found from [28]. The outcomes of the GP model were analyzed by using the statistical indexes and compared with the experimental results.

**Table 2.** Parameters of the genetic programming (GP) models.

| Parameter | Definition | Value (Model 1) | Value (Model 2) | Value (Model 3) |
|---|---|---|---|---|
| 1 | Function set | +, −, *, √, ^2, cos, sin, exp | +, −, *, √, ^2, cos, sin, exp | +, −, *, √, ^2, cos, sin, exp |
| 2-1 | The terminal set for $\tau_b/\bar{\tau}_b$ | b/B, B/H, Fr, Re | b/B, B/H, Fr | b/B, B/H |
| 2-2 | The terminal set for $\tau_w/\bar{\tau}_w$ | z/H, B/H, Fr, Re | z/H, B/H, Fr | z/H, B/H |

### 2.4. Adaptive Neuro Fuzzy Inference System (ANFIS)

ANFIS is designed to provide the requisite inputs and outputs for adaptive networks to build fuzzy rules with acceptable membership functions. ANFIS is a common and cardinal programming method that uses fuzzy theory to write fuzzy if-then rules and fuzzy logic bases that map from a given input information to the desired output. An adaptive network is a multilayer feed-forward artificial neural network (ANN) with; partially or entirely adaptive nodes in which the outputs are predicted on adaptive node parameters and the parameter adjustment is specified by the learning rules due to the error term. In adaptive ANFIS, hybrid learning is generally a learning form [29].

*2.5. Criteria for Statistical Assessment*

Maximum error (ME), mean absolute error (MAE), root mean square error (RMSE) and Nash–Sutcliffe efficiency (NSE) are the four statistical evaluation parameters used to determine the Tsallis entropy, GP model and ANFIS model performance, which are measured as follows [30,31].

$$\text{ME} = \text{Max}|P_i - O_i| \tag{21}$$

$$\text{MAE} = \frac{1}{N}\sum_{i=1}^{N}|P_i - O_i| \tag{22}$$

$$\text{RMSE} = \sqrt{\frac{\sum_{i=1}^{n}(P_i - O_i)^2}{n}} \tag{23}$$

$$\text{NSE} = 1 - \frac{\sum_{i=1}^{n}(P_i - O_i)^2}{\sum_{i=1}^{n}(O_i - \bar{O})^2} \tag{24}$$

where $O_i$ is the observed parameter value, $P_i$ predicted parameter value, $\bar{O}$ is the mean value observed parameter value and $n$ number of samples.

## 3. Results

*3.1. Modeling of GP*

In this section, sensitivity of the GP model for any input parameter is evaluated by adding all four inputs to the models first. Each parameter is then omitted and a total of three separate versions are checked. The GP models used for data on the bed and wall are described as:

For the bed

$$\text{GP Model}(1): \frac{b}{B}, \frac{B}{H}, \text{Fr}, \text{Re}$$

$$\text{GP Model}(2): \frac{b}{B}, \frac{B}{H}, \text{Fr},$$

$$\text{GP Model}(3): \frac{b}{B}, \frac{B}{H}$$

For the wall:

$$\text{GP Model}(1): \frac{z}{H}, \frac{B}{H}, \text{Fr}, \text{Re}$$

$$\text{GP Model}(2): \frac{z}{H}, \frac{B}{H}, \text{Fr}$$

$$\text{GP Model}(3): \frac{z}{H}, \frac{B}{H}$$

For each channel section, three different models were evaluated to investigate the effect of each input parameter in the GP modeling. The findings of the modeling of bed shear stress show that the GP model (1) had the lowest error consisting of input parameters (*b/B*, *B/H*, Fr and Re). The results of the modeling of bed shear stress revealed that the lowest error (average RMSE = 0.0874) was observed in the GP model (1) consisting of input parameters (*b/B*, *B/H*, Fr and Re) and modeled wall shear stress, the GP model (1) had the lowest input error (*z/H*, *B/H*, Fr and Re) (average RMSE = 0.0692), so that the *B/H* had a major influence on the GP model and validated the effects of model (1). By performing a sensitivity analysis, since the flow situation was fully developed, the Reynolds number could be ignored and the parameter was eliminated in model (2).

As shown in Table 3, by omitting the Reynolds number (Re) in the input parameters, there was no significant difference. On the other hand, because all the experiments examined the subcritical flow conditions, the effect Froude number could be ignored and the parameter was eliminated in model 3. By eliminating the Reynolds number and Froude number parameters, the GP model performance did not change much, and the GP model could be deduced to be insensitive to the *B*/*H* parameter. The *B*/*H* ratio was obviously important in the estimation of shear stress, as this parameter played a significant role in the equations stated. Therefore, the model 3 for the bed and wall was chosen as the most suitable model. The results of the most accurate GP model and experimental bed and wall data are shown in the form of the scatter plots in Figures 3 and 4. As seen in statistical analysis, the GP model outcomes were very similar to the bed and wall shear stress line fitted. Dimensionless bed shear stress modeling with GP was superior to dimensionless wall shear stress modeling with average NSE of 0.945 and 0.8266, respectively, and both models were superior to the other GP models in this study. In order to decide the best answer, the best feedback should be treated as a pattern. Different important parameters in modeling, such as population members, number of generations, tree structures size, etc., should be carefully determined in the first step with regard to the consumer of the data examined.

The scale of each configuration of the tree will play a major role in the final model's accuracy. Determining the numbers greater than the optimal value reduced the precision of the test results and it prevented displaying the models, which are not presented largely because the models generated by genetic programming were of a very long-scale in order to measure the shear stress. The method of fitting models resulting from genetic programming against experimental results of parameters 2.86, 4.51, 7.14 and 13.95 are shown in Figure 4. The statistical analysis results of GP model predictions tabulated in Table 3.

**Table 3.** Performance metric of GP models to predict SSD.

| B/H | Input Variable | Bed | | | | Input Variable | Wall | | | |
|---|---|---|---|---|---|---|---|---|---|---|
| | | ME | MAE | RMSE | NSE | | ME | MAE | RMSE | NSE |
| 2.86 | b/B, B/H | 0.2259 | 0.0713 | 0.1051 | 0.9382 | z/H, B/H | 0.0728 | 0.0217 | 0.0870 | 0.7277 |
| 2.86 | b/B, B/H, Fr | 0.2445 | 0.1038 | 0.1206 | 0.9456 | z/H, B/H, Fr | 0.0693 | 0.0257 | 0.0821 | 0.7759 |
| 2.86 | b/B, B/H, Fr, Re | 0.2338 | 0.0837 | 0.1062 | 0.947 | z/H, B/H, Fr, Re | 0.0363 | 0.0617 | 0.0516 | 0.8021 |
| 4.51 | b/B, B/H | 0.1450 | 0.0962 | 0.0995 | 0.9889 | z/H, B/H | 0.0874 | 0.0530 | 0.0972 | 0.8987 |
| 4.51 | b/B, B/H, Fr | 0.1019 | 0.0642 | 0.0638 | 0.9903 | z/H, B/H, Fr | 0.0818 | 0.0302 | 0.0890 | 0.8972 |
| 4.51 | b/B, B/H, Fr, Re | 0.0927 | 0.0473 | 0.0526 | 0.9911 | z/H, B/H, Fr, Re | 0.0546 | 0.0202 | 0.0701 | 0.8548 |
| 7.14 | b/B, B/H | 0.0826 | 0.0348 | 0.0468 | 0.9955 | z/H, B/H | 0.0589 | 0.1153 | 0.0648 | 0.9049 |
| 7.14 | b/B, B/H, Fr | 0.0851 | 0.0408 | 0.0493 | 0.9962 | z/H, B/H, Fr | 0.0330 | 0.0321 | 0.0617 | 0.8566 |
| 7.14 | b/B, B/H, Fr, Re | 0.0889 | 0.0466 | 0.0533 | 0.9958 | z/H, B/H, Fr, Re | 0.0422 | 0.0424 | 0.0507 | 0.8982 |
| 13.95 | b/B, B/H | 0.1619 | 0.0908 | 0.1059 | 0.8534 | z/H, B/H | 0.0716 | 0.0926 | 0.1126 | 0.7758 |
| 13.95 | b/B, B/H, Fr | 0.2678 | 0.1398 | 0.1566 | 0.8511 | z/H, B/H, Fr | 0.0559 | 0.0264 | 0.1117 | 0.8097 |
| 13.95 | b/B, B/H, Fr, Re | 0.2005 | 0.1269 | 0.1376 | 0.8667 | z/H, B/H, Fr, Re | 0.0612 | 0.0720 | 0.1045 | 0.7916 |
| 13.95 | b/B, B/H | 0.2259 | 0.0713 | 0.1051 | 0.9382 | z/H, B/H | 0.0728 | 0.0217 | 0.0870 | 0.7277 |

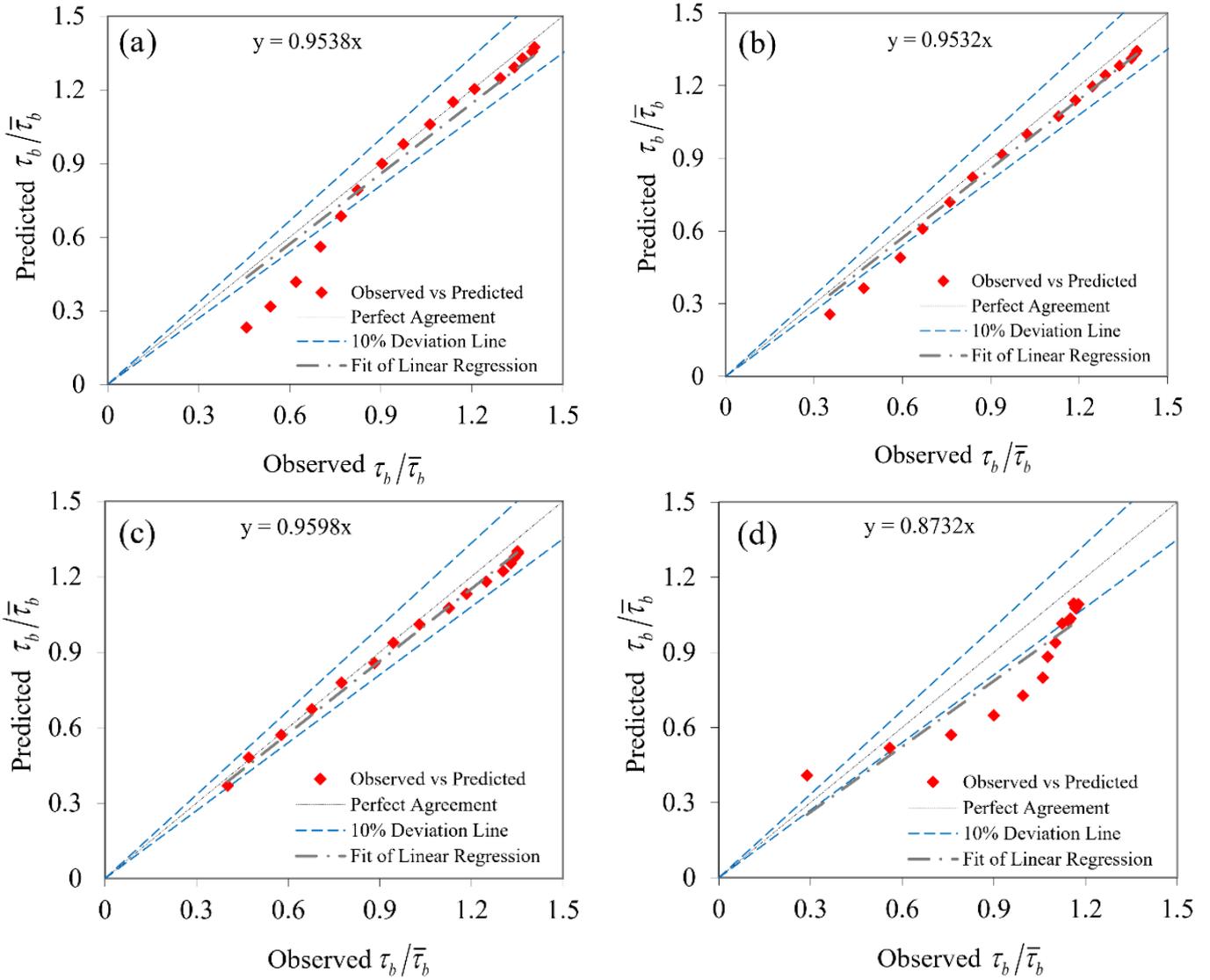

**Figure 3.** Comparison to the estimate of $\tau_b/\bar{\tau}_b$ between the observed and predicted GP for (**a**) *B*/*H* = 2.86, (**b**) *B*/*H* = 4.51, (**c**) *B*/*H* = 7.14 and (**d**) *B*/*H* = 13.95.

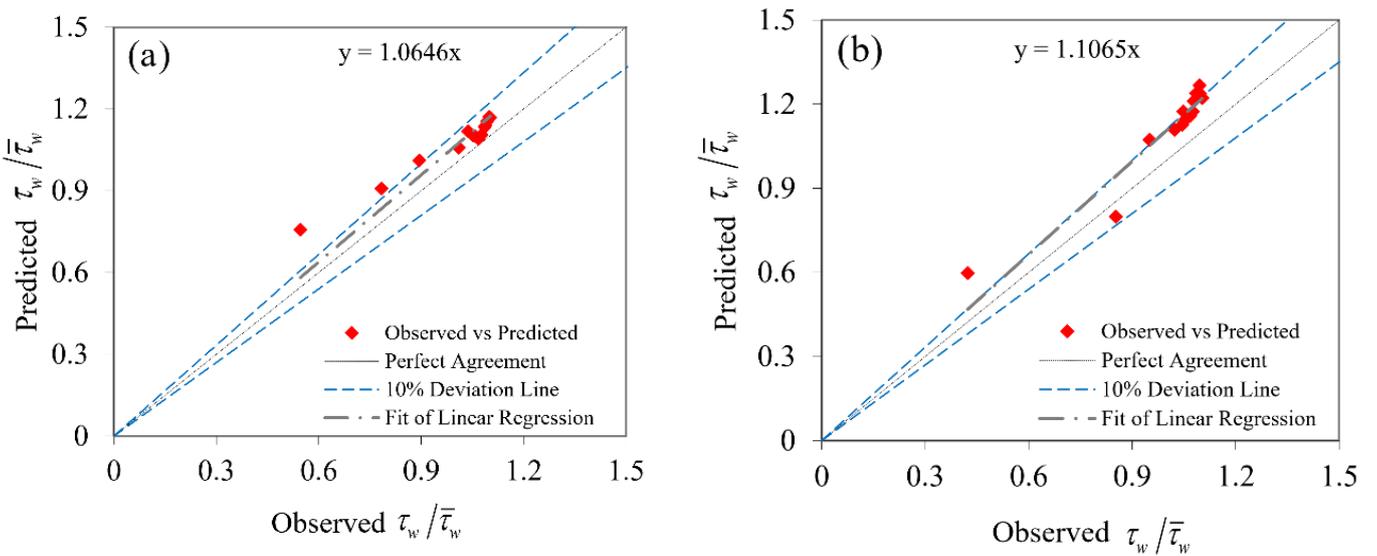

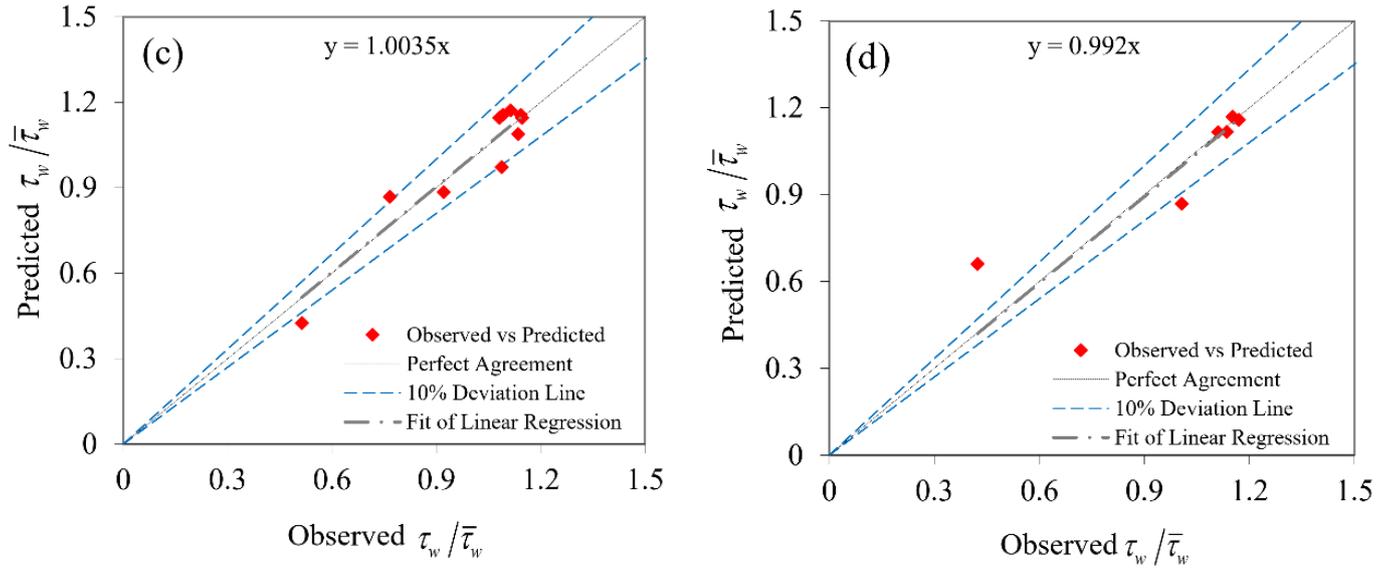

**Figure 4.** Comparison to the estimate of $\tau_w/\bar{\tau}_w$ between the observed and predicted GP for (**a**) *B/H* = 2.86, (**b**) *B/H* = 4.51, (**c**) *B/H* = 7.14 and (**d**) *B/H* = 13.95.

*3.2. ANFIS Modeling*

For this purpose, 70% of the experimental data was used for network training and the remaining 30% was used for testing results. As input parameters to the model, the parameters *b/B* and *B/H* for bed and *z/H* and *B/H* for the wall were presented. Figure 5 shows the performance of the ANFIS model to estimate the bed SSD ($\tau_b$) and Figure 6 shows the performance of the ANFIS model to estimate the wall SSD ($\tau_w$), 30% of the data, which were not used in the training stage would be used to evaluate the performance of the model. The results of statistical indexes for modeling shear stress with ANFIS are summarized in Table 4. As well, the estimating bands of the four above parameters using to determine the shear stress are shown in Figure 5. Skewness results obtained from statistical prediction dimensionless parameters.

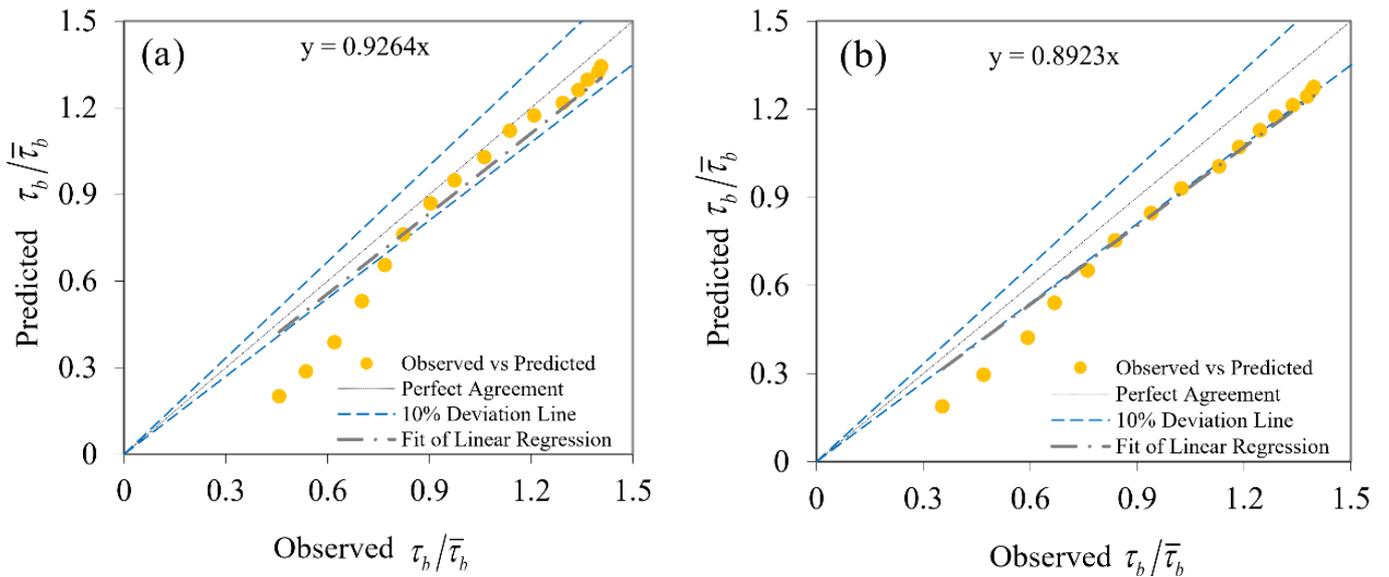

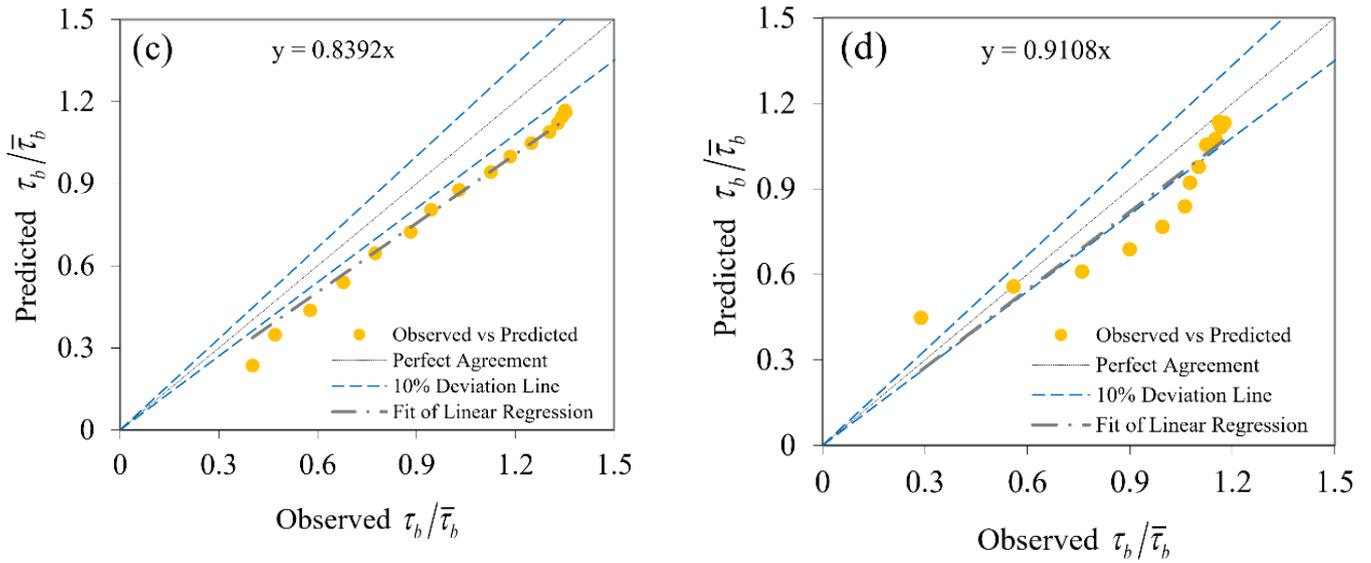

**Figure 5.** Comparison to the estimate of $\tau_b/\overline{\tau}_b$ $\frac{\tau_b}{\overline{\tau}}$ between the observed and predicted adaptive neuro-fuzzy inference system (ANFIS) for (**a**) *B/H* = 2.86, (**b**) *B/H* = 4.51, (**c**) *B/H* = 7.14, and (**d**) *B/H* = 13.95.

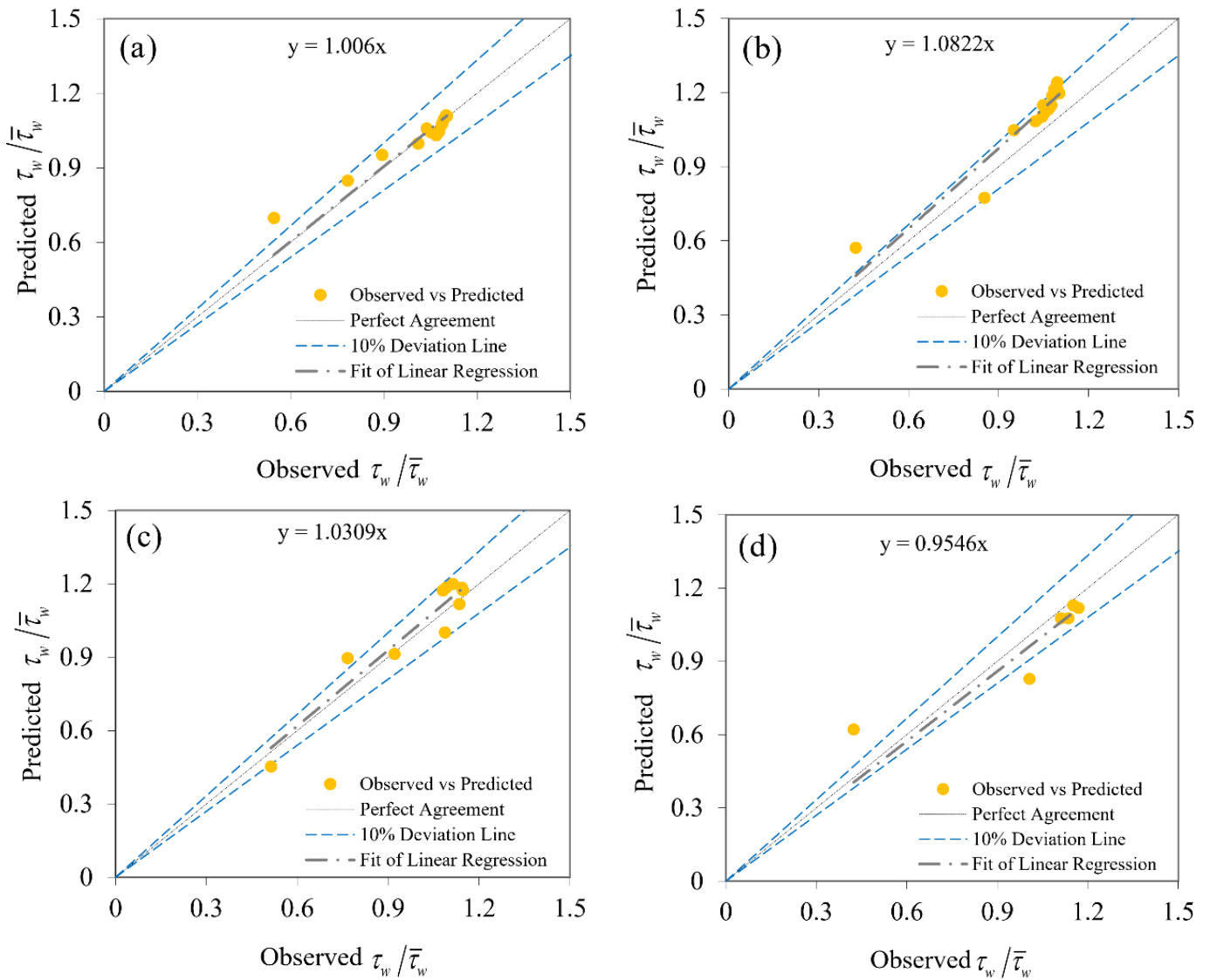

**Figure 6.** Comparison to the estimate of $\tau_w/\overline{\tau}_w$ between the observed and predicted ANFIS for (**a**) *B/H* = 2.86, (**b**) *B/H* = 4.51, (**c**) *B/H* = 7.14, and (**d**) *B/H* = 13.95.

**Table 4.** Performance metric of the ANFIS model to predict SSD.

| B/H | Bed | | | | Wall | | | |
|---|---|---|---|---|---|---|---|---|
| | ME | MAE | RMSE | NSE | ME | MAE | RMSE | NSE |
| 2.86 | 0.2559 | 0.0991 | 0.1268 | 0.9279 | 0.0383 | 0.0314 | 0.0492 | 0.8026 |
| 4.51 | 0.1728 | 0.1240 | 0.1266 | 0.9744 | 0.0870 | 0.0959 | 0.1004 | 0.9033 |
| 7.14 | 0.2157 | 0.1699 | 0.1724 | 0.9871 | 0.0868 | 0.0634 | 0.0745 | 0.907 |
| 13.95 | 0.2278 | 0.1048 | 0.1271 | 0.8482 | 0.1792 | 0.0909 | 0.1145 | 0.7752 |

*3.3. Comparison of the GP Model, Tsallis Entropy and ANFIS*

The results of the best GP models and Tsallis entropy in shear stress prediction were compared with the experimental results of Lashkar-Ara and Fatahi [8] in this section. Figures 7 and 8 show the experimental results and SSD predictions with different models in a smooth rectangular channel for *B/H* equal to 2.85, 4.51, 7.14 and 13.95. Additionally, the performance metric of the shear stress estimate by the Tsallis entropy model is shown in Table 5. As shown in these statistics, all of the test evidence used to model the SSD using the GP was is realized. For the training stage for modeling SSD in the rectangular channel using the GP model, 70 percent of all data were used, and 30 percent of the data were used for the testing process. As shown in Figure 7, for *B/H*= 2.86, 4.51, 7.14 and 13.95, the GP model predicted the bed shear stress better than the Tsallis entropy model. In Figure 8c,d, for *B/H* = 7.14 and 13.95, the GP model predicted wall shear stress better than the Tsallis entropy model, but in Figure 8a,b, the Tsallis entropy was more accurately modeled to predict wall shear stress than the GP model. Additionally, the GP model estimated bed and wall shear stress better than the Tsallis entropy-based model at rising flow depth. It is understandable that the channel architecture was challenging when a model expected higher shear stress values. It is therefore not cost-effective to use the Tsallis entropy method. When the GP model's observations were more accurate, it could be used to design stable channels more consistently. The GP model estimated the bed shear better than the ANFIS model for *B/H*= 2.86, 4.51, 7.14 and 13.95. For *B/H* = 2.86, the ANFIS model estimated the shear stress better than the GP model, but the GP model estimated the wall shear stress better than the ANFIS model in *B/H* = 4.51, 7.14 and 13.95. The GP model demonstrated superior efficiency to the Tsallis entropy-based model, while both models neglected the influence of secondary flows. It can be inferred that the GP model of bed and wall shear stress estimation was more sensitive than the Tsallis entropy method overestimated the values of bed shear stress and the GP model's outcomes were greater. The bed shear stress values decreased at the middle of the channel (Figure 7), which varied from other situations. From Figures 7 and 8, it can be shown that the GP model's fit line was similar to the 45-degree line than the other ones, and with a higher value of NSE, its predictions were more reliable. In predicting the position of maximal shear stress, both the GP and Tsallis-entropy based models displayed the same pattern as the centerline of the channel, which was consistent with the experimental outputs.

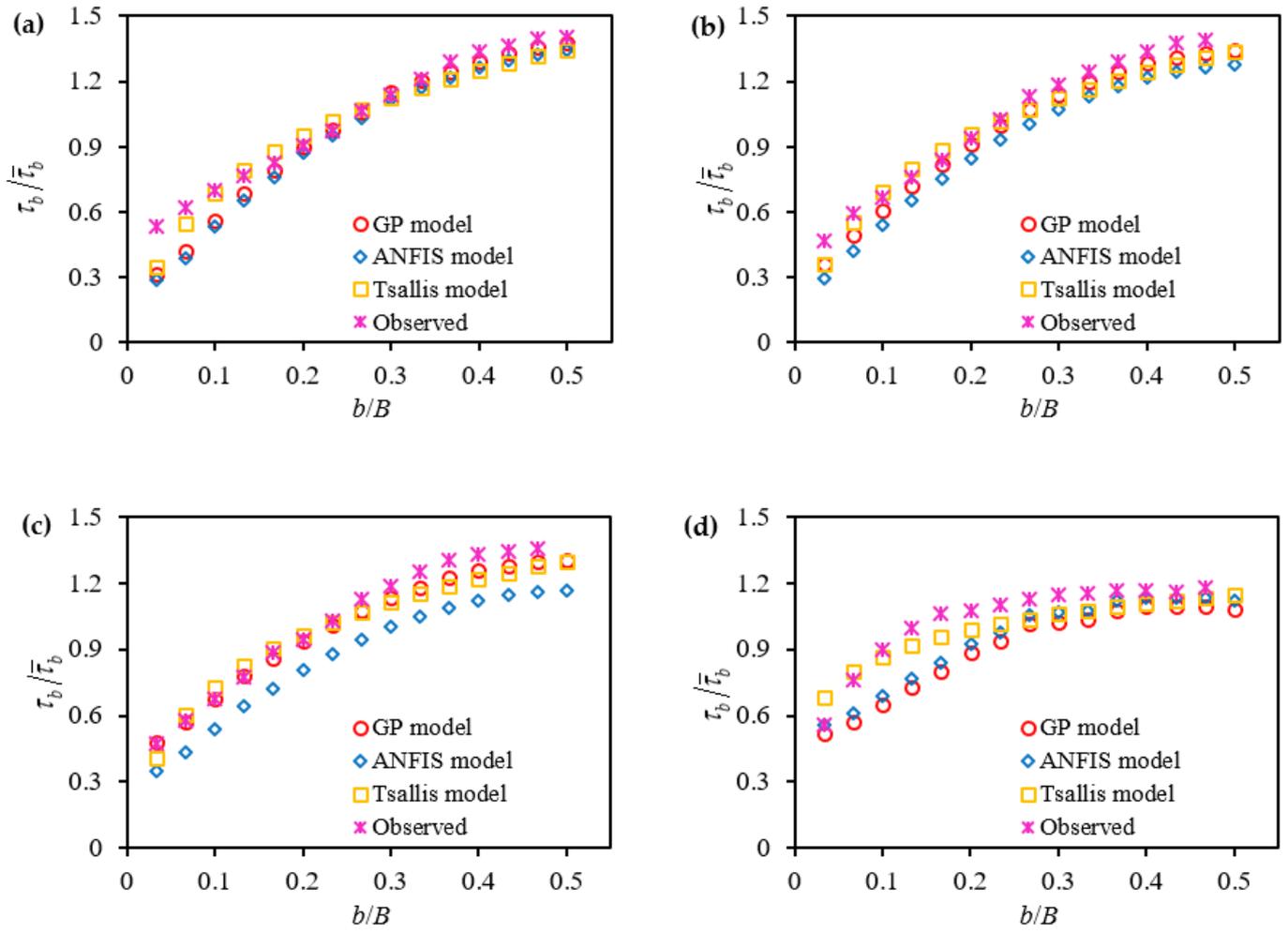

**Figure 7.** The dimensionless bed shear stress distribution for (**a**) $B/H$ = 2.86, (**b**) $B/H$ = 4.51, (**c**) $B/H$ = 7.14 and (**d**) $B/H$ = 13.95.

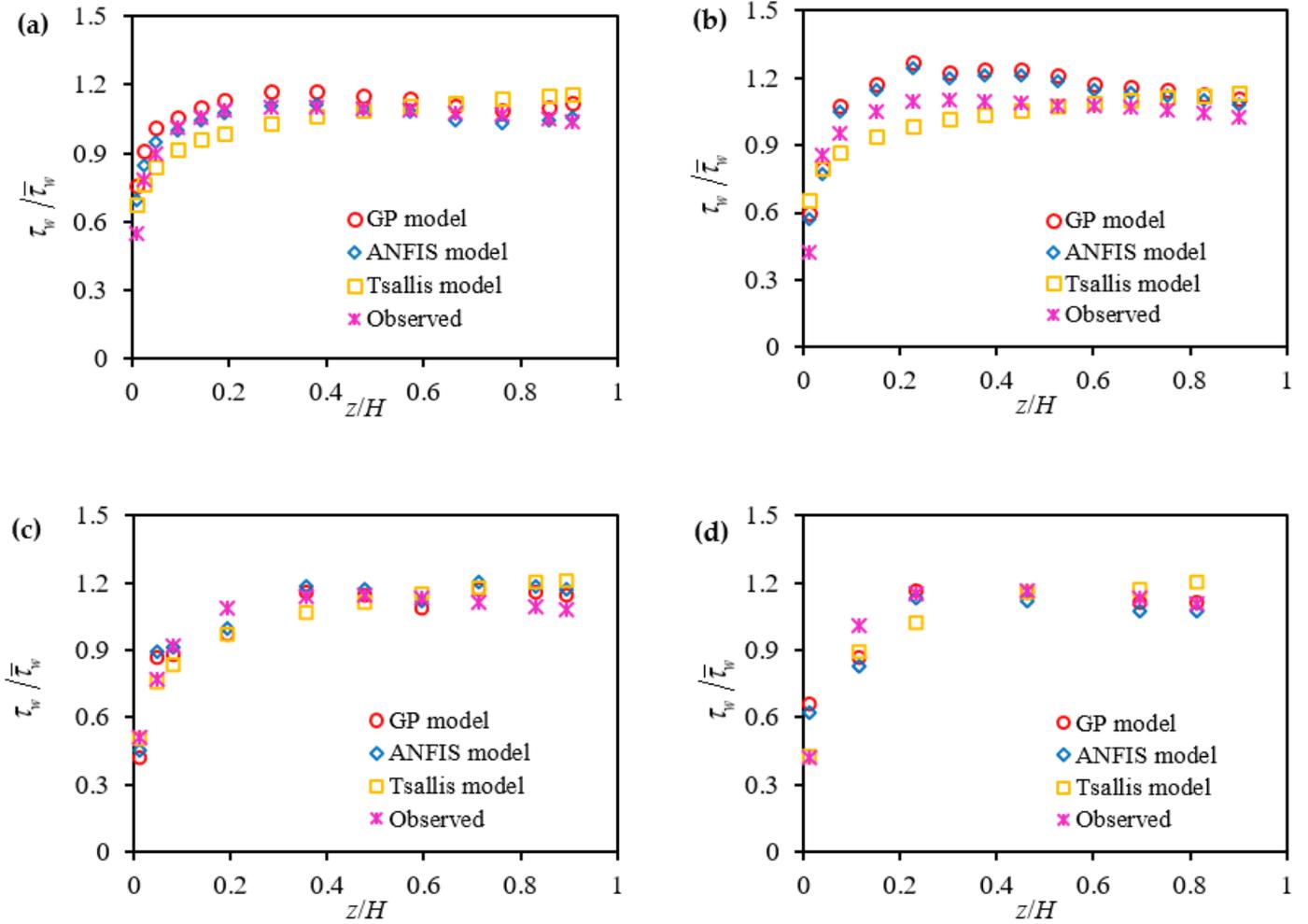

**Figure 8.** The dimensionless wall shear stress distribution for (**a**) *B/H* = 2.86, (**b**) *B/H* = 4.51, (**c**) *B/H* = 7.14 and (**d**) *B/H* = 13.95.

**Table 5.** Performance metric of Tsallis entropy to predict SSD.

| B/H | Bed | | | | Wall | | | |
|---|---|---|---|---|---|---|---|---|
| | ME | MAE | RMSE | NSE | ME | MAE | RMSE | NSE |
| 2.86 | 1.252 | 0.0531 | 0.0706 | 0.9276 | 1.3145 | 0.0622 | 0.0797 | 0.7721 |
| 4.51 | 1.476 | 0.0522 | 0.0625 | 0.9425 | 1.3741 | 0.0749 | 0.0894 | 0.7632 |
| 7.14 | 1.538 | 0.0672 | 0.0685 | 0.9310 | 1.6254 | 0.0631 | 0.0738 | 0.8275 |
| 13.95 | 1.511 | 0.0643 | 0.0840 | 0.8426 | 1.2562 | 0.0893 | 0.1094 | 0.8398 |

## 4. Conclusions

The wall and bed shear stresses in a smooth rectangular channel measured experimentally for different aspect ratios. Two soft computing models GP and ANFIS proposed to estimate SSD in rectangular channel. In addition, the results of GP and ANFIS model compared with a Tsallis based equation. Our research had some main findings as follows:

1. The effect of different input variable on the result was investigated to find the best input combination.
2. In the present study *B/H* had the highest effect on the prediction power.
3. For bed shear stress predictions, the GP model, with an average RMSE of 0.0893 performed better than the Tsallis entropy-based equation and ANFIS model with RMSE of 0.0714 and 0.138 respectively.

4. To estimate the wall shear stress distribution the proposed ANFIS model, with an average RMSE of 0.0846 outperformed the Tsallis entropy-based equation with an RMSE of 0.0880 followed by the GP model with an RMSE of 0.0904.

Our finding suggests that the proposed GP algorithm could be used as a reliable and cost-effective algorithm to enhance SSD prediction in rectangular channels.

**Author Contributions:** Supervision, B.L.-A.; Writing – original draft, B.L. -A. and N.K.; Writing – review & editing, Z.S.K. and A.M. All authors have read and agreed to the published version of the manuscript.

**Funding:** Not applicable.

**Institutional Review Board Statement:** Not applicable.

**Informed Consent Statement:** Not applicable.

**Data Availability Statement:** Not applicable.

**Conflicts of Interest:** The authors declare no conflict of interest.